\documentclass[twoside,twocolumn,9pt]{article}
\usepackage{extsizes}
\usepackage[super,sort&compress,comma]{natbib} 
\usepackage[version=3]{mhchem}
\usepackage[left=1.5cm, right=1.5cm, top=1.785cm, bottom=2.0cm]{geometry}
\usepackage{balance}
\usepackage{mathptmx,mathtools}
\usepackage{sectsty}
\usepackage{graphicx} 
\usepackage{lastpage}
\usepackage[format=plain,justification=justified,singlelinecheck=false,font={stretch=1.125,small,sf},labelfont=bf,labelsep=space]{caption}
\usepackage{float}
\usepackage{fancyhdr}
\usepackage{fnpos}
\usepackage[english]{babel}
\addto{\captionsenglish}{%
  
}
\usepackage{array}
\usepackage{droidsans}
\usepackage{charter}
\usepackage[T1]{fontenc}
\usepackage[usenames,dvipsnames]{xcolor}
\usepackage{setspace}
\usepackage[compact]{titlesec}
\usepackage{hyperref}

\usepackage{epstopdf}

\definecolor{cream}{RGB}{222,217,201}

\begin{document}

\pagestyle{fancy}
\thispagestyle{plain}
\fancypagestyle{plain}{
\renewcommand{\headrulewidth}{0pt}
}

\makeFNbottom
\makeatletter
\renewcommand\LARGE{\@setfontsize\LARGE{15pt}{17}}
\renewcommand\Large{\@setfontsize\Large{12pt}{14}}
\renewcommand\large{\@setfontsize\large{10pt}{12}}
\renewcommand\footnotesize{\@setfontsize\footnotesize{7pt}{10}}
\makeatother

\renewcommand{\thefootnote}{\fnsymbol{footnote}}
\renewcommand\footnoterule{\vspace*{1pt}%
\color{cream}\hrule width 3.5in height 0.4pt \color{black}\vspace*{5pt}} 
\setcounter{secnumdepth}{5}

\makeatletter 
\renewcommand\@biblabel[1]{#1}            
\renewcommand\@makefntext[1]%
{\noindent\makebox[0pt][r]{\@thefnmark\,}#1}
\makeatother 
\renewcommand{\figurename}{\small{Fig.}~}
\sectionfont{\sffamily\Large}
\subsectionfont{\normalsize}
\subsubsectionfont{\bf}
\setstretch{1.125} 
\setlength{\skip\footins}{0.8cm}
\setlength{\footnotesep}{0.25cm}
\setlength{\jot}{10pt}
\titlespacing*{\section}{0pt}{4pt}{4pt}
\titlespacing*{\subsection}{0pt}{15pt}{1pt}

\fancyfoot{}
\fancyfoot[LO,RE]{\vspace{-7.1pt}\includegraphics[height=9pt]{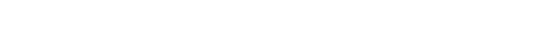}}
\fancyfoot[CO]{\vspace{-7.1pt}\hspace{13.2cm}\includegraphics{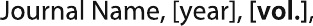}}
\fancyfoot[CE]{\vspace{-7.2pt}\hspace{-14.2cm}\includegraphics{RF}}
\fancyfoot[RO]{\footnotesize{\sffamily{1--\pageref{LastPage} ~\textbar  \hspace{2pt}\thepage}}}
\fancyfoot[LE]{\footnotesize{\sffamily{\thepage~\textbar\hspace{3.45cm} 1--\pageref{LastPage}}}}
\fancyhead{}
\renewcommand{\headrulewidth}{0pt} 
\renewcommand{\footrulewidth}{0pt}
\setlength{\arrayrulewidth}{1pt}
\setlength{\columnsep}{6.5mm}
\setlength\bibsep{1pt}

\makeatletter 
\newlength{\figrulesep} 
\setlength{\figrulesep}{0.5\textfloatsep} 

\newcommand{\topfigrule}{\vspace*{-1pt}%
\noindent{\color{cream}\rule[-\figrulesep]{\columnwidth}{1.5pt}} }

\newcommand{\botfigrule}{\vspace*{-2pt}%
\noindent{\color{cream}\rule[\figrulesep]{\columnwidth}{1.5pt}} }

\newcommand{\dblfigrule}{\vspace*{-1pt}%
\noindent{\color{cream}\rule[-\figrulesep]{\textwidth}{1.5pt}} }

\makeatother

\twocolumn[
  \begin{@twocolumnfalse}
{\includegraphics[height=30pt]{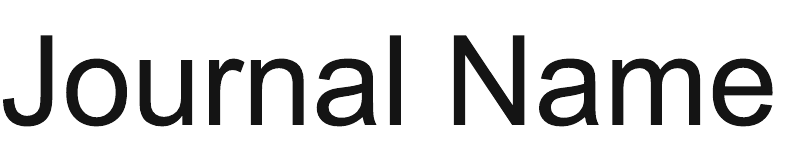}\hfill\raisebox{0pt}[0pt][0pt]{\includegraphics[height=55pt]{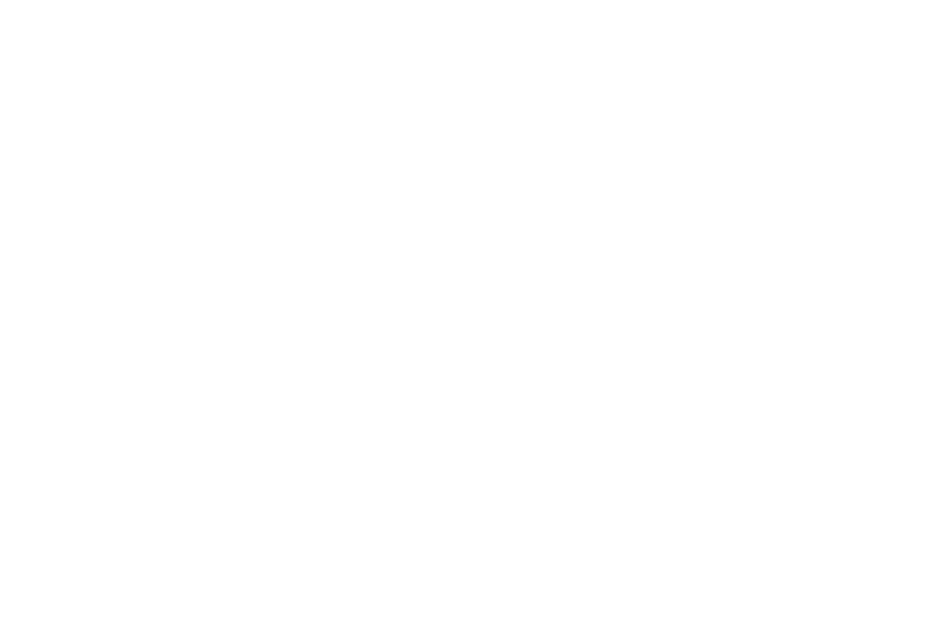}}\\[1ex]
\includegraphics[width=18.5cm]{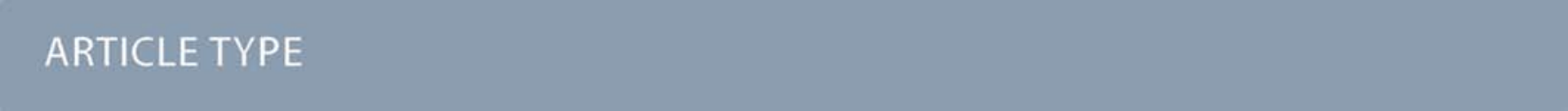}}\par
\vspace{1em}
\sffamily
\begin{tabular}{m{4.5cm} p{13.5cm} }

\includegraphics{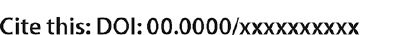} & \noindent\LARGE{\textbf{CO$_2$-driven diffusiophoresis and water cleaning: similarity solutions for predicting the exclusion zone in a channel flow$^\dag$}} \\
\vspace{0.3cm} & \vspace{0.3cm} \\

 & \noindent\large{Suin Shim,\textit{$^{a,\ast}$} Mrudhula Baskaran,\textit{$^{b}$} Ethan H. Thai,\textit{$^{c}$} and Howard A. Stone\textit{$^{a,\ddag}$}} \\
 \vspace{-0.3cm}

\includegraphics{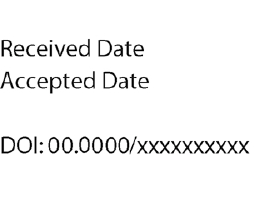} & \noindent\normalsize{We investigate experimentally and theoretically diffusiophoretic separation of negatively charged particles in a rectangular channel flow, driven by CO$_2$ dissolution from one side-wall. Since the negatively charged particles create an exclusion zone near the boundary where CO$_2$ is introduced, we model the problem by applying a shear flow approximation in a two-dimensional configuration. From the form of the equations we define a similarity variable to transform the reaction-diffusion equations for CO$_2$ and ions and the advection-diffusion equation for the particle distribution to ordinary differential equations. The definition of the similarity variable suggests a characteristic length scale for the particle exclusion zone. We consider height-averaged flow behaviors in rectangular channels to rationalize and connect our experimental observations with the model, by calculating the wall shear rate as functions of channel dimensions. Our observations and the theoretical model provide the design parameters such as flow speed, channel dimensions and CO$_2$ pressure for the in-flow water cleaning systems.} \\

\end{tabular}

 \end{@twocolumnfalse} \vspace{0.7cm}

  ]

\renewcommand*\rmdefault{bch}\normalfont\upshape
\rmfamily
\section*{}
\vspace{-1cm}


\footnotetext{\textit{$^{a}$Department of Mechanical and Aerospace Engineering, Princeton University, Princeton, NJ 08544, USA }}
\footnotetext{\textit{$^{b}$Unsteady Flow Diagnostics Laboratory, Institute of Mechanical Engineering, École Polytechnique Fédérale de Lausanne, 1015, Lausanne, Switzerland}}
\footnotetext{\textit{$^{c}$Department of Electrical Engineering, Princeton University, Princeton, NJ 08544, USA }}

\footnotetext{\dag~Electronic Supplementary Information (ESI) available: [details of any supplementary information available should be included here]. See DOI: 00.0000/00000000.}

\footnotetext{$\ast$ sshim@princeton.edu\\ \ddag hastone@princeton.edu}

\section{Introduction}

Diffusiophoresis is the spontaneous motion of colloidal particles under a solute concentration gradient. Since the first recognition \cite{der1,der2}, the phenomenon has been widely studied theoretically and experimentally \cite{dukhin1, prieve1, prieve2, prieve3, prieve4, abecassis, velegol1, ez1, palacci1, velegol2, size, squires1, ault1, gupta1, battat1, wilson1, gupta2}. When the solute is an electrolyte, the net motion of particles has electrophoretic and chemiphoretic contributions, induced by, respectively, the diffusivity difference(s) among the ions and the osmotic pressure gradient near the particle surface \cite{prieve2}.

Various experimental investigations have been reported for the diffusiophoresis of micron-sized particles, either in the presence or absence of liquid flow \cite{velegol2, size, shin1,battat1, wilson1, gupta2, squires2, lee}. Recently, it has been demonstrated that dissolution of gas can drive diffusiophoresis of charged particles \cite{shim,shin1,pulse,hsc,shin2,orest}. In particular, CO$_2$ was used in several earlier studies due to its ubiquity and importance in many applications. From a transport perspective the most important feature of CO$_2$ as a means for driving diffusiophoresis is that it undergoes rapid dissolution and reaction in water, creating H$^+$ and HCO$_3^-$ ions by the dissociation of H$_2$CO$_3$. These ions have a large difference in their diffusivities with $D_{\text{H}^+} = 9.3 \times 10^{-9}$ m$^2$s$^{-1}$ and $D_{\text{HCO}_3^-} = 1.2 \times 10^{-9}$ m$^2$s$^{-1}$, leading to a large diffusion potential and diffusiophoretic mobility of suspended charged particles \cite{shim,shin1,pulse,hsc,shin2,orest}.

As the nature of diffusiophoresis allows moving charged particles with an imposed directionality, it has been suggested for water cleaning. In our previous work \cite{hsc}, we demonstrated in a Hele-Shaw geometry that bacterial cells can be excluded from a region near a CO$_2$ source by CO$_2$-driven diffusiophoresis. Shin \textit{et al.}\cite{shin1}, Lee \textit{et al.}\cite{lee}, and Seo \textit{et al}\cite{seo}, demonstrated removing particles from water by diffusiophoretic exclusion of particles, using a CO$_2$ gas chamber or Nafion as the sources of ions, respectively, in a configuration where the solute gradient is created perpendicular to the liquid flow. The main mechanism for such a configuration is that either a particle free zone or an accumulation zone forms near the walls depending on the diffusiophoretic mobility and the direction of the concentration gradient. However, the reported experiments use one or two geometrical conditions (in each study), and we sought to understand what could be achieved in larger scales with the analogous exclusion zone formation. 

\begin{figure*}[t]
\centering
\includegraphics[width=7.2in]{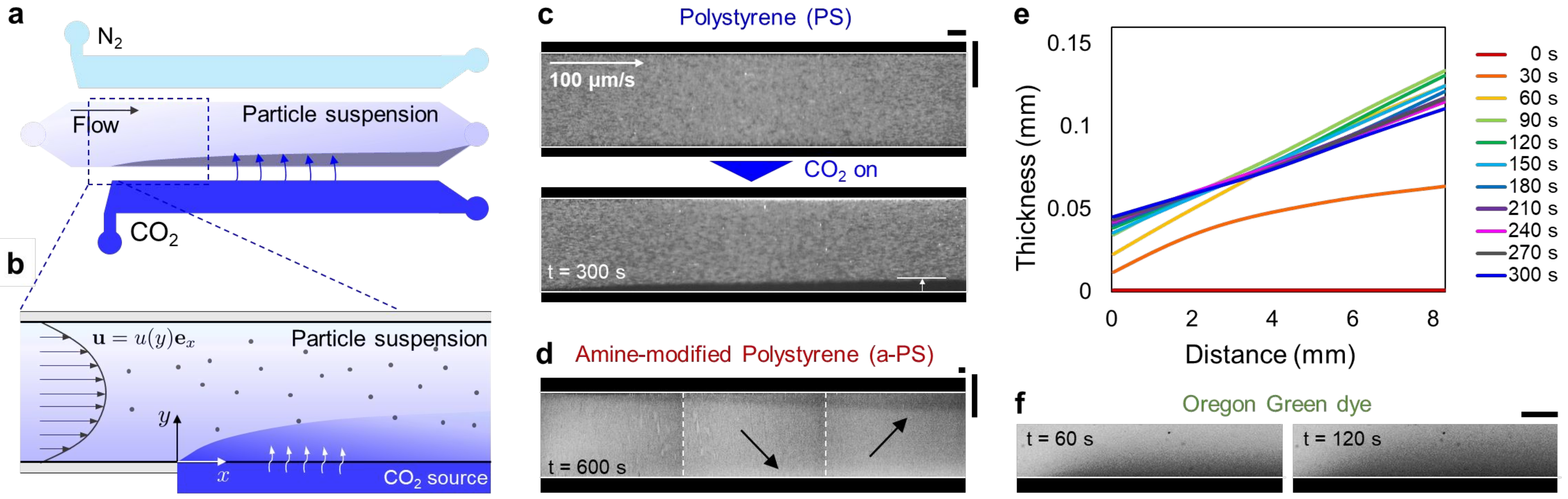}
\caption{\label{fig1} CO$_2$-driven diffusiophoresis in the channel flow. (a,b) Schematic of the experimental system. (a) Schematic of the microfluidic channel design. (b) Zoomed-in view of the channel flow. We define the $x$ coordinate along the channel starting from the position of the CO$_2$ inlet. (c) Exclusion zone formation in the flow of negatively charged polystyrene (PS) particles. There is a finite thickness exclusion zone forming at the lower boundary where CO$_2$ diffuses in. (d) Diffusiophoresis of positively charged amine-modified polystyrene (a-PS) particles. There is an accumulation of a-PS particles at the lower (CO$_2$ entering) boundary. Near the top boundary, the particle concentration decreases, but this is not an exclusion (zero-concentration) zone. (c,d) The average flow speed is $\langle u \rangle=100~\mu$m/s. (e) Time evolution of the exclusion zone in an 1-mm-width channel experiment with the flow speed $\langle u \rangle = 100~\mu$m/s. (f) Visualization of CO$_2$ diffusion using a pH sensitive dye (Oregon Green 488; Thermo Fisher Scientific). Scale bars are 500 $\mu$m. }
\end{figure*}

To better understand the transport process in such in-flow water cleaning systems, and to discuss scale-up possibilities, mathematical models are needed. A simple increase in the geometry cannot be applied for effective scale-up of the system due to the nonlinearity of the diffusiophoresis, possible transitions from laminar to turbulent flows, and other effects (e.g. gravity) that influence the system as the length scale increases. For example, CO$_2$ dissolved water is denser than normal water, so a simple increase in the length scale will cause buoyancy effects that complicate the flow and the motion of the particles. In such case a different design is necessary.


In this paper, we are interested in a channel flow configuration where the driving source of diffusiophoretic water cleaning is dissolved CO$_2$, and the concentration gradient is perpendicular to the flow \cite{shin1}. Through the detailed investigation of the steady-state flow and the particle exclusion zone using both theory and experiments, we identify important parameters that directly affect the scalability of the system and the exclusion zone. In particular, we apply a shear flow approximation to the near-wall region in the channel flow, and develop similarity solutions for the distribution of ions and particles. Then with control experiments and numerical calculations considering the effect of channel geometry we identify parameters that influences the size of the exclusion zone, which directly affects the efficiency of the water cleaning. Our goal is not to provide an optimal solution or the most efficient channel design. Rather, we try to sort out the important system parameters to apply to the device design process.

\smallskip
\section{Formation of the particle exclusion zone in channel flow}

We create a concentration gradient of CO$_2$ in a channel flow by placing a CO$_2$ channel and the N$_2$ channel adjacent to the liquid channel. The permeable PDMS membranes between the channels (thickness $= 300~\mu$m) let the gases dissolve into the liquid channel (see Figure \ref{fig1}(a)). As a result, a steady state concentration gradient of CO$_2$ is set up in the liquid channel as described in Figure \ref{fig1}(a,b). In the liquid channel, a particle suspension (polystyrene; 0.03 vol$\%$ in DI water) flows into the channel at a constant mean flow velocity. As a result of the CO$_2$ concentration gradient which leads to the ion (H$^+$ and HCO$_3^-$) concentration gradients across the channel, the charged particles either form an exclusion zone or accumulate near the CO$_2$-side boundary depending on the surface charge of the particles. Negatively charged polystyrene (PS) particles form a particle-free zone near the CO$_2$-side wall as shown in Figure \ref{fig1}(c) while positively charged amine-modified polystyrene (a-PS) particles accumulate (Figure \ref{fig1}(d)). Details of the experimental setup is described in the Materials and Methods section. In this study, we focus on the particle exclusion zone formation near the CO$_2$-side boundary. Figure \ref{fig1}(e) shows the measured shape of the particle exclusion zone at different times when negatively charged particles flow into the channel at a mean speed $\langle u \rangle = 100~\mu$m/s. The boundary of the exclusion zone is plotted versus distance along the channel, which is measured from the CO$_2$ inlet position (Figure \ref{fig1}(a,b)). We note that some time after the CO$_2$ stream is turned on, the particle exclusion zone grows then reach a steady-state as shown in Figure \ref{fig1}(e) (a longer time experiment is presented in the SI). The nonzero thickness near $x=0$ is due to the CO$_2$ diffusion in the PDMS walls, which makes the influx of CO$_2$ occur from further upstream. The diffusion of CO$_2$ in the liquid channel can be visualized with a pH-sensitive dye (Oregon Green 488, Figure \ref{fig1}(f)). 

The diffusiophoretic velocity $\mathbf{u}_p$ of particles under the ion concentration gradient $\nabla c_i$ is written as
\begin{equation}
\mathbf{u}_p = \Gamma_p \nabla \ln c_i = \Gamma_p \frac{\nabla c_i}{c_i}~ \label{u_p},
\end{equation}
where $\Gamma_p$ is the diffusiophoretic mobility of the particles under a concentration gradient of a z:z electrolyte \cite{prieve2};
\begin{equation}
\Gamma_p =  \frac{\epsilon}{ \mu} \frac{k T}{ z e} \left[ \beta \zeta - \frac{2 k T}{z e} \text{ln} \left(1- \text{tanh}^2 \frac{z e \zeta}{4 k T} \right) \right]~.
\end{equation}

\begin{figure*}[t!]
\centering
\includegraphics[width=4.65in]{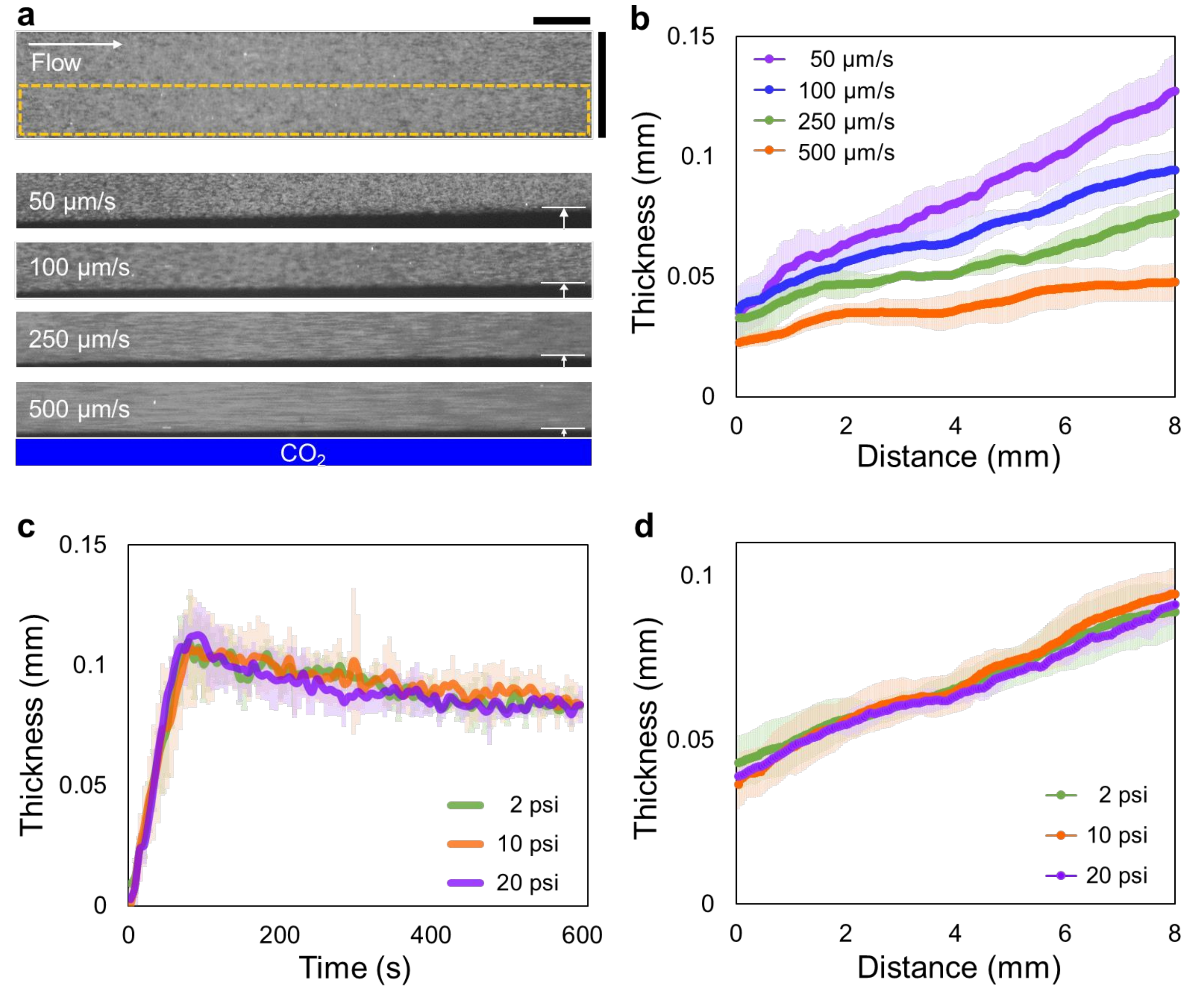}
\caption{\label{fig2} Effect of flow speed and the CO$_2$ pressure on the exclusion zone formation. (a) Experimental images taken in the 1-mm-width channel at different flow speeds. For slower particle flow, the exclusion zone thickness is larger. (b) Measured thickness of exclusion zones obtained at different flow speeds. (c,d) Effect of CO$_2$ pressure on the exclusion zone formation. Mean flow velocity is 100 $\mu$m/s. (c) Time evolution of the exclusion zone thickness at $x=8$ mm for different CO$_2$ pressures. (d) Exclusion zone profile at $t=300$ s, obtained at different CO$_2$ pressures. Scale bars are 500 $\mu$m. }
\end{figure*}
Here $\epsilon$, $\mu$, $k$, $T$, $e$ and $\zeta$ are, respectively, the dielectric permittivity of the solution, dynamic viscosity of the solution, Boltzmann's constant, absolute temperature and the zeta potential of the particle. $\displaystyle{\beta = \frac{D_+-D_-}{D_+ +D_-}}$ is the diffusivity difference factor, which determines the strength of the local electric field induced by the difference in diffusivities of the ions.  

To obtain initial understanding of the system, we controlled the flow rate in the liquid channel and varied the CO$_2$ pressure. 
In the channel flow, the shape of steady-state exclusion zone can be varied by changing the flow speed (Figure \ref{fig2}(a,b)). For faster flows, we obtain thinner exclusion zones simply because the advection and diffusion of CO$_2$ and ions are perpendicular to each other, and the diffusiophoresis has diffusive feature (note that the diffusiophoretic mobility has a dimension $\ell^2/t$). 

In the experiments where CO$_2$ pressure was varied, we observed that there was not much change in the shape of the exclusion zone. For three different CO$_2$ pressures (2, 10 and 20 psi) applied to the CO$_2$ stream, both time evolution (Figure \ref{fig2}(c)) and the steady state shape (measured at $t=300$ s, Figure \ref{fig2}(d)) of the exclusion zone were not affected. We interpret that this no-CO$_2$-pressure-dependence is due to the fast reaction and the logarithmic feature of diffusiophoresis. From equation (\ref{u_p}), we can estimate the particle velocity by an order of magnitude estimate $\displaystyle{u_p \approx \frac{\Gamma_p}{c_i} \frac{c_i}{w} = \frac{\Gamma_p}{w}}$, where $w$ is the width of the channel over which the ion concentration gradient is formed. This supports our observation that CO$_2$ pressure did not affect the shape of particle exclusion zone. The scaling analysis will not apply to situations where the ion concentration is either very small or very large, which is considered outside the range of our experiments.

Our initial investigation of the particle exclusion zone with various flow speeds and CO$_2$ pressures tells us that having slower flow is good for having a larger exclusion zone, and we only need a moderate amount of CO$_2$ to achieve such an exclusion zone. However, this is not enough information for scale-up of the system. For example, slower flow will result in a smaller amount of clean liquid collected at the outlet, which lowers the system efficiency. Therefore, we investigate the system in more detail with a mathematical model.

\begin{figure*}[t]
\centering
\includegraphics[width=7in]{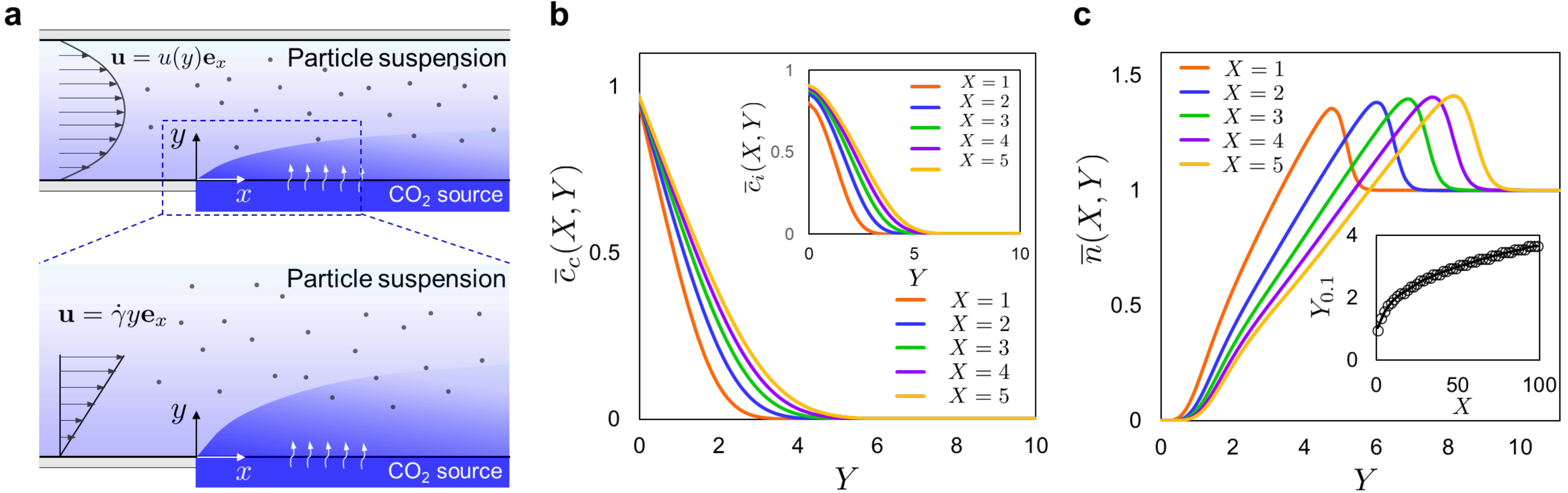}
\caption{\label{fig3} Two dimensional model calculation with shear flow approximation. (a) Schematic of shear flow approximation near the CO$_2$ entering boundary. For $Pe \gg 1$, the boundary layer approximation can be applied for the diffusion terms, and we can solve a steady-state, scale invariant problem using the shear flow approximation. (b) Solutions for equations (8) and (9) for nondimensional CO$_2$ and ion concentrations. Along the flow -- increasing X --, both CO$_2$ and ions diffuse into the flow in $Y$ direction. Note that the flux of ions is zero at the CO$_2$-side boundary. (c) Nondimensional particle distribution along the flow (solutions to equation (12)). Particles are excluded from the CO$_2$ entering boundary ($Y=0$), and the local accumulation of particles propagate downstream. (c) Inset: An exclusion zone profile (arbitrarily defined for the value of $Y$ where $\overline{n}(X,Y) \approx 0.1$) was obtained from the nondimensional particle distribution and plotted versus $X$. This provides rough estimate of particle free zone profile. }
\end{figure*}

\section{Steady-state, 2D model with a shear flow approximation}

Consider a fully developed flow of a particle suspension in a rectangular microfluidic channel. If CO$_2$ dissolves into the liquid from one wall as decribed in Fig.~\ref{fig3}(a), we expect to achieve diffusiophoresis of particles induced by an ion concentration gradient  in the channel created by the reaction \cite{shim,shin1,hsc,pulse,shin2,orest}
\begin{equation}
\text{CO$_2$} + \text{H$_2$O} \xrightleftharpoons[k_r]{k_f} \text{H$^+$} + \text{HCO$_3^-$}~.
\end{equation}
 Assume a 2D pressure driven flow with flow velocity \textbf{u}$=u(y)$ \textbf{e$_x$}, then the reaction-diffusion equation for dissolved CO$_2$ and dissociated ions can be written as \cite{hsc,orest}
\begin{align}
\frac{\partial c_c}{\partial t} + u\frac{\partial c_c}{\partial x} &= D_c \left(\frac{\partial^2 c_c}{\partial x^2}+\frac{\partial^2 c_c}{\partial y^2}\right) - (k_f c_c - k_r c_i^2),\\
\frac{\partial c_i}{\partial t} + u\frac{\partial c_i}{\partial x} &= D_A \left(\frac{\partial^2 c_i}{\partial x^2}+\frac{\partial^2 c_i}{\partial y^2}\right)  +(k_f c_c - k_r c_i^2),
\end{align}
where $c_c$, $c_i$ are concentrations of CO$_2$ and ions, and $D_c$ and $D_A$ are, respectively, diffusivity of CO$_2$ ($D_c = 1.9 \times 10^{-9}$ m$^2$/s) and the ambipolar diffusivity of the ions. $k_f = 0.039$ s$^{-1}$ and $k_r = 9.2 \times 10^4$ L/mol$\cdot$s are forward and reverse reaction rate constants \cite{yaws, haynes, jolly}. The equations can be solved numerically with appropriate boundary and initial conditions for $u(y)$. For long times we neglect the time derivative to obtain the steady state concentration field of the ions.

Here we want to seek for solutions for a simplified problem within a small length scale. Consider CO$_2$ dissolution and dissociation within $x$ small enough that the concentration distribution of dissolved CO$_2$ does not extend across the entire channel width. Then the region of interest is a thin layer near the wall where CO$_2$ is supplied to the liquid phase. Then we can assume that the liquid flow is simply a shear flow with velocity \textbf{u}$= \dot{\gamma} y$ \textbf{e$_x$} where $\dot{\gamma}$ is shear rate (Fig.~\ref{fig3}(a)-bottom panel), and apply the boundary layer approximation to further calculations, which is well-known as the L\'ev\^eque approximation in the Chemical Engineering literature. \cite{ristenpart}. 

Now consider steady state reaction-diffusion equations for CO$_2$ and ions in this near-wall region;
\begin{align}
\dot{\gamma}y\frac{\partial c_c}{\partial x} &= D_c \frac{\partial^2 c_c}{\partial y^2}- (k_f c_c - k_r c_i^2)~,\label{shearcc}\\
\dot{\gamma}y\frac{\partial c_i}{\partial x} &= D_A \frac{\partial^2 c_i}{\partial y^2}+ (k_f c_c - k_r c_i^2)~. \label{shearci} 
\end{align}
Boundary conditions are, $c_c(x,0)=c_c^{\text{sat}}$, $c_c(0,y)=c_c(x,\infty)=0$, $\displaystyle{\frac{\partial c_i}{\partial y}\bigg |_{(x,0)} =0}$ and $c_i(0,y)=c_i(x,\infty)=0$, where $c_c^{\text{sat}}$ is the equilibrium concentration of CO$_2$ under the applied CO$_2$ pressure $p_c$ so that $c_c^{\text{sat}} = k_h p_c$ ($k_h$ is Henry's Law constant). Note that the ions satisfy a zero flux condition at the CO$_2$-side boundary, and also note that under chemical equilibrium $k_f c_c^{\text{sat}} = k_r (c_i^{\text{sat}})^2$, where $c_i^{\text{sat}}$ is the saturation concentration of ions under the applied gas pressure $p_c$. Equations (\ref{shearcc}) and (\ref{shearci}) can be nondimensionalized with $\overline{c}_c=c_c / c_c^{\text{sat~}}$, $\overline{c}_i=c_i / c_i^{\text{sat~}}$, $X=x/\sqrt{D_c / \dot{\gamma}}$ and $Y=y/\sqrt{D_c / \dot{\gamma}}$. 
The nondimensional diffusion-reaction equations are
\begin{align}
Y \frac{\partial \overline{c}_c}{\partial X} &= \frac{\partial^2 \overline{c}_c}{\partial Y^2} - \frac{k_f}{\dot{\gamma}}(\overline{c}_c-\overline{c}_i^2)~, \label{shearndcc}\\
Y \frac{\partial \overline{c}_i}{\partial X}&= \bar{D}_A \frac{\partial^2 \overline{c}_i}{\partial Y^2} + \frac{k_r c_i^{\text{sat}}}{\dot{\gamma}}(\overline{c}_c-\overline{c}_i^2)~, \label{shearndci}
\end{align}
where $\bar{D}_A = D_A/D_c$ is the nondimensional ambipolar diffusivity. The nondimensional boundary conditions are then, $\overline{c}_c(X,0)=1$, $\overline{c}_c (0,Y) = \overline{c}_c (X,\infty) = 0$, $\displaystyle{\frac{\partial \overline{c}_i}{ \partial Y} \bigg|_{(X,0)} = 0}$ and $\overline{c}_i (0,Y) = \overline{c}_i (X,\infty) = 0$. We note that $k_f/\dot\gamma$ is dimensionless, and for $\dot\gamma=2$ s$^{-1}$, $k_f/\dot\gamma \approx 0.02 \ll 1$. Equations (\ref{shearndcc}) and (\ref{shearndci}) were solved with Matlab and plotted versus $Y$ for different $X$ values in Figure \ref{fig3}(b). The plots show that both CO$_2$ and ions diffuse into the channel (toward larger $Y$) as the liquid flows downstream (increasing $X$). Note that the CO$_2$ has a constant pressure (and so the concentration) condition at the boundary and the solutions for ions show zero-flux condition at the boundary. Using the concentration of ions and the diffusiophoretic velocity of particles, now we can solve for the particle distribution in the channel also under the shear flow approximation.

The 2D unsteady advection-diffusion equation for particles of concentration $n$ is 
\begin{equation}
\frac{\partial n}{\partial t} + u\frac{\partial n}{\partial x}+\frac{\partial (u_{p, x} n)}{\partial x} + \frac{\partial (u_{p, y}  n)}{\partial y} = D_p \left(\frac{\partial^2 n}{\partial x^2}+\frac{\partial^2 n}{\partial y^2}\right)~,
\end{equation}
where $u_{p,x}$ and $u_{p,y}$ are, respectively, the $x$ and $y$ components of diffusiophoretic velocity $u_p$. For small $x$ and $y$, we can apply the shear flow approximation to the particle equation. Note that under the boundary layer approximation, we can neglect the diffusiophoresis in the $x$ direction for $\displaystyle{\frac{\Gamma_p}{x} < \dot{\gamma} y ~\Rightarrow~ \frac{\Gamma_p}{\dot{\gamma}(D_c /{\dot{\gamma}})}=\frac{\Gamma_p}{D_c} <1}$. This is valid for most experimental situations so we keep the condition for further calculations. Then we obtain a steady-state equation for particles,
\begin{equation}
\dot\gamma y  \frac{\partial n}{\partial x} + \Gamma_p  \frac{\partial }{\partial y} \left(n\frac{\partial \ln c_i}{\partial y}\right) = D_p \frac{\partial^2 n}{\partial y^2}~.
\end{equation}
Boundary conditions are, $n(0,y)=n(X,\infty)=n_0$ and $\displaystyle{-\Gamma_p \frac{\partial \ln c_i}{\partial y} n + D_p \frac{\partial n}{\partial y} = 0}$ at $y=0$, where $n_0$ is the unperturbed particle concentration.

The equation can be nondimensionalized with $\overline{n}=n/n_0$, $X=x/\sqrt{D_c / \dot{\gamma}}$ and $Y=y/\sqrt{D_c / \dot{\gamma}}$, and thus
\begin{equation}
Y \frac{\partial \overline{n}}{\partial X} + \bar{\Gamma}_p \frac{\partial}{\partial Y} \left( \overline{n} ~ \frac{\partial \ln \overline{ c}_i }{\partial Y} \right) = \bar{D}_p \frac{\partial^2 \overline{n}}{\partial Y^2}~, \label{shearndn}
\end{equation}
where $\bar{\Gamma}_p = \Gamma_p/D_c$ and $\bar{D}_p = D_p/D_c$. Equation (\ref{shearndn}) is solved and the solution is plotted versus $Y$ for different positions along the channel in Figure \ref{fig3}(c). Note that the solutions show zero particle concentration region near $Y=0$ which indicates the particle exclusion zone. We calculated the value of $Y$ where $\overline{n}(X,Y) = 0.1$, then plotted this location versus $X$ (inset of Figure \ref{fig3}(c)). This plot shows an approximate shape of the steady-state particle exclusion zone obtained from the model calculation. For the shear rate $\dot\gamma = 5$ s$^{-1}$, the nondimensional thickness of exclusion zone $Y_{0.1} \approx 4$ yields a dimensional length scale is $y_{0.1} \approx 80~\mu$m, which provides a reasonable estimate of a typical length scale of exclusion zone thickness. 

Performing numerical calculation of the partial differential equations (PDEs) provides a good estimate of the particle exclusion zone, but we can further simplify our equations. From the form of the equation, we can deduce a similarity variable and analyze the system with a set of ordinary differential equations (ODEs).

\smallskip
\section{Similarity solutions for predicting particle exclusion zone}

\begin{figure*}[t!]
\centering
\includegraphics[width=4.8in]{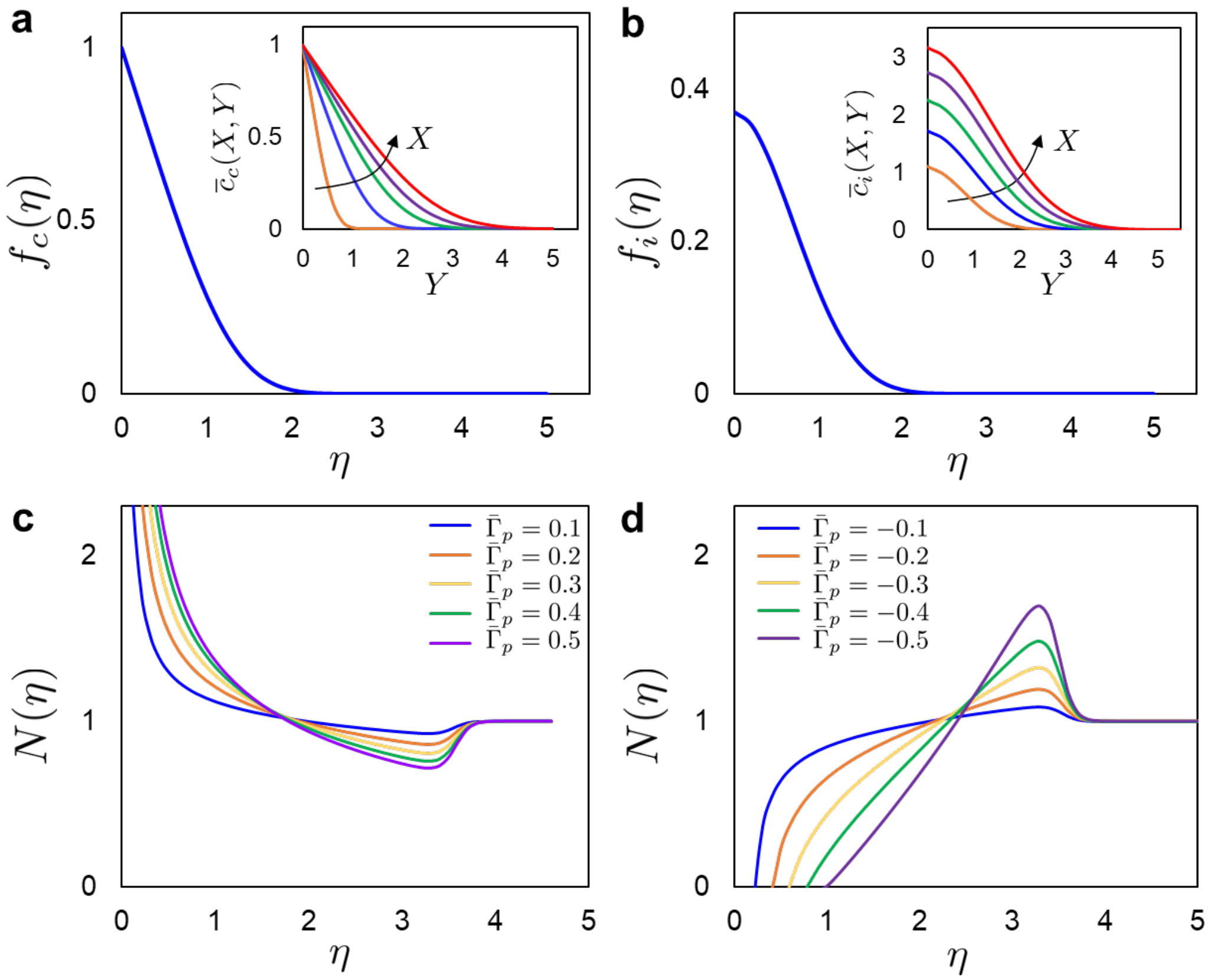}
\caption{\label{fig4} Similarity solutions for concentrations of the CO$_2$, ions and the particles. (a) Similarity solution for CO$_2$ concentration plotted versus $\eta$. Inset: CO$_2$ concentration plotted versus $Y$ for different values of $X$. (b) Ion concentration $f_i(\eta)$ plotted versus $\eta$. Inset: ion concentration $\overline{c}_i$ is plotted versus $Y$ for different $X$'s. Note that the inset graphs in (a) and (b) are solutions to approximated equations, and so they show different solutions from Figure \ref{fig3}. (c, d) Solutions of equation (eq) are plotted for different values of $\bar{\Gamma}_p$. (c) Positively charged particles ($\bar{\Gamma}_p > 0$) tend to accumulate more as the mobility increases near $\eta \rightarrow 0$. (d) Negatively charged particles ($\bar{\Gamma}_p > 0$) move away from the boundary ($\eta = 0$), and the larger the mobility is, more particles are cleared away from the CO$_2$ boundary. There is a finite value of $\eta$ for each negative $\bar{\Gamma}_p$ which indicate the thickness of particle free zone in terms of similarity variable.}
\end{figure*}

From the structure of equation (\ref{shearndcc}), we can consider a similarity variable $\displaystyle{\eta = \frac{Y}{g(X)}}$ and find a similarity solution $f_c(\eta) = \overline{c}_c (X,Y)$. Substituting $\overline{c}_c$ with $f_c(\eta)$ in equation (\ref{shearndcc}), we get
\begin{equation}
-\eta^2 \frac{d f_c}{d \eta} g' g^2 = \frac{d^2 f_c}{d \eta^2} -\frac{k_f g^2}{\dot\gamma} (f_c - \overline{c}_i^2)~.\label{subfceta}
\end{equation}
Finding $g(X)$ that satisfies $g' g^2 = 1$ gives $g(X)=(3 X)^{\frac{1}{3}}$, and thus equation (\ref{subfceta}) becomes  
\begin{equation}
-\eta^2 \frac{d f_c}{d \eta} = \frac{d^2 f_c}{d \eta^2} - \frac{k_f}{\dot\gamma} (f_c - \overline{c}_i^2) (3 X)^{\frac{2}{3}}~.
\end{equation}
Since $\displaystyle{\frac{k_f}{\dot\gamma} X^{\frac{2}{3}} \ll 1}$ for small $X$ (note that $k_f/\dot\gamma \approx 0.02$ for $\dot\gamma = 2$ s$^{-1}$), the reaction term can be neglected. Therefore, we solve an ODE 
\begin{equation}
-\eta^2 \frac{d f_c}{d \eta} = \frac{d^2 f_c}{d \eta^2}~, \label{eqnfceta}
\end{equation} 
with the boundary conditions $f_c(0)=1$ and $f_c(\infty) = 0$. This is a similarity transform of the well-known L\'ev\^eque problem, and the solution to equation (\ref{eqnfceta}) is 
\begin{equation}
f_c(\eta) = \frac{\int_\eta^\infty e^{-s^3/3} ds}{\int_0^\infty e^{-s^3/3} ds} = \frac{\Gamma \left(\frac{1}{3},\frac{\eta^3}{3}\right)}{\Gamma \left(\frac{1}{3}\right)}~.\label{fcetasol}
\end{equation}
Here $\Gamma(\cdot)$ is the Gamma function and $\Gamma(\cdot,\cdot)$ is the incomplete Gamma function, i.e. $\displaystyle{\Gamma \left( \frac{1}{3}, \frac{\eta^3}{3}\right) = \int_{\frac{\eta^3}{3}}^\infty t^{-\frac{2}{3}} e^{-t} dt}$. The solution (\ref{fcetasol}) is plotted in Figure \ref{fig4}(a). The inset of Figure \ref{fig4}(a) shows the solutions to the L\'ev\^eque problem plotted versus $Y$ at different values of $X$. The plots show the diffusion of CO$_2$ into the channel starting from the boundary ($\eta=Y=0$).

The same analysis can be applied to the concentration of ions near the boundary. For equation (\ref{shearndci}), we can find a similarity solution $f_i(\eta)$ by defining $\displaystyle{\overline{c}_i (X,Y) = \frac{k_r c_i^{\text{sat}}}{\dot\gamma} (3 X)^{\frac{2}{3}} f_i(\eta)}$ with $\displaystyle{\eta=\frac{Y}{(3X)^{\frac{1}{3}}}}$. Substituting this into equation (\ref{shearndci}), we obtain
\begin{equation}
2 \eta f_i -\eta^2 \frac{d f_i}{d \eta} = \bar{D}_A \frac{d^2 f_i}{d \eta^2} + \left( f_c -\left( \frac{k_r c_i^{\text{sat}}}{\dot\gamma} \right)^2 (3 X)^{\frac{4}{3}} f_i^2 \right)~,
\end{equation}
where the term $\displaystyle{\left( \frac{k_r c_i^{\text{sat}}}{\dot\gamma} \right)^2 (3 X)^{\frac{4}{3}} \ll 1}$ for small $X$. Therefore,
\begin{equation}
2 \eta f_i - \eta^2 \frac{d f_i}{d \eta} = \bar{D}_A \frac{d^2 f_i}{d \eta^2} + f_c~, \label{eqnfieta}
\end{equation}
and the boundary conditions are $f_i'(0)=0$ and $f_i(\infty)=0$. The ion concentration satisfies no-flux condition at the CO$_2$-side boundary. 
This equation shows that the ion concentration depends on the CO$_2$ concentration by the forcing term $f_c$. Equation (\ref{eqnfieta}) is solved and plotted in Figure \ref{fig4}(b). The inset of Figure \ref{fig4}(b) is the plot of $\displaystyle{\overline{c}_i (X,Y) = \frac{k_r c_i^{\text{sat}}}{\dot\gamma} (3 X)^{\frac{2}{3}} f_i(\eta)}$ versus $Y$ at different values of $X$. For increasing $X$, the ions created by the reaction diffuse into the channel. 

Finally, we can consider the particle distribution near the CO$_2$ boundary where CO$_2$ enters. Define $N(\eta)=\overline{n}(X,Y)$ and substitute into equation (\ref{shearndn}) we obtain
\begin{equation}
-\eta^2 \frac{d N}{d \eta} + \bar{\Gamma}_p \frac{d N}{d \eta} \frac{d \ln f_i}{d \eta} = \bar{D}_p \frac{ d^2 N}{d \eta^2} - \bar{\Gamma}_p N \frac{ d^2 \ln f_i}{d \eta^2}~. \label{eqnNeta}
\end{equation}
The boundary conditions are, $\displaystyle{-\bar{\Gamma}_p N \frac{d \ln f_i}{d \eta}+ \bar{D}_p \frac{d N}{d \eta} = 0}$ at $\eta = 0$ and $N(\infty)=1$. Under typical experimental conditions, $\bar{D}_p \ll \bar{\Gamma}_p$, and thus we can neglect the particle diffusion. As a result, we solve a first-order ODE for $N(\eta)$, 
\begin{equation}
-\eta^2 \frac{d N}{d \eta} + \bar{\Gamma}_p \frac{d}{d\eta}\left( N \frac{d \ln f_i}{d \eta}\right) = 0~,\label{Netanodiff}
\end{equation}
with a boundary condition $N(\infty)=1$. Both for positively charged ($\bar\Gamma_p > 0$) and negatively charged particles ($\bar\Gamma_p < 0$), the equation (\ref{Netanodiff}) is solved and plotted versus $\eta$ in Figure \ref{fig4}(c,d) with five different magnitudes of the mobility $\bar{\Gamma}p$. Solutions to the full equation (\ref{eqnNeta}) are plotted in the SI. For positively charged particles (Figure \ref{fig4}(c)), there is accumulation (increase in particle concentration) near the boundary $\eta=0$, and the particle accumulation is enhanced for larger mobilities. For negatively charged particles (Figure \ref{fig4}(d)), we note that for each value of diffusiophoretic mobility, there is a critical value of $\eta$ below which the particle concentrations are effectively zero. We call this cutoff value $\eta_c$. In our experiments, the diffusiophoretic mobility of the negatively charged polystyrene particles corresponds to $\bar{\Gamma}_p=-0.4$, and the critical value $\eta_c = 0.78$.

\begin{figure}[b!]
\centering
\includegraphics[width=2.76in]{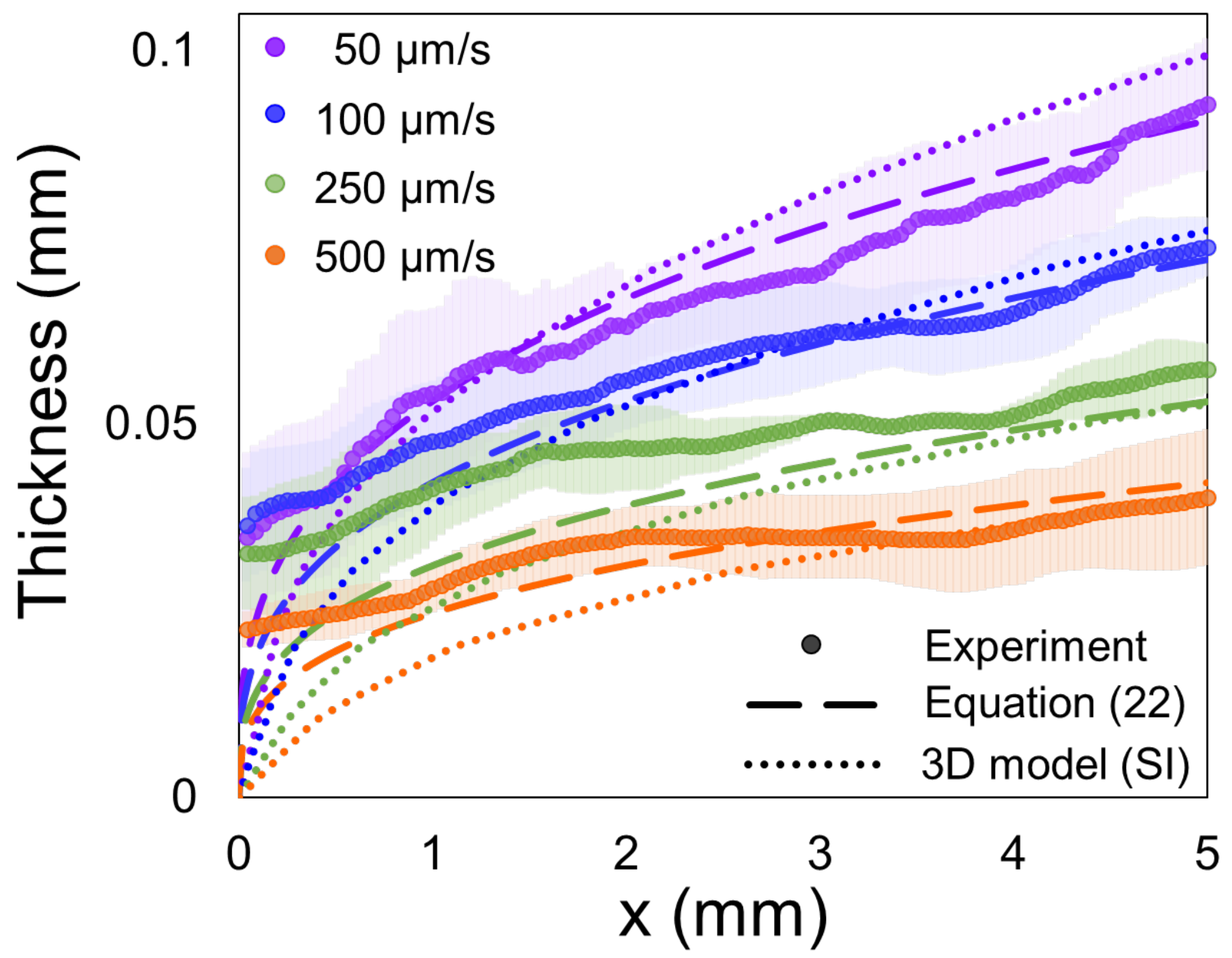}
\caption{\label{fig4-5} Comparison between experiments and equation (\ref{eq22}) for different flow speeds. Experiments are done in a 1-mm-width and a 100-micron-height channel. With the choice of $\alpha=0.45$, equation (\ref{eq22}) fits the experimental measurements under different conditions.}
\end{figure}

\begin{figure*}[t!]
\centering
\includegraphics[width=7.3in]{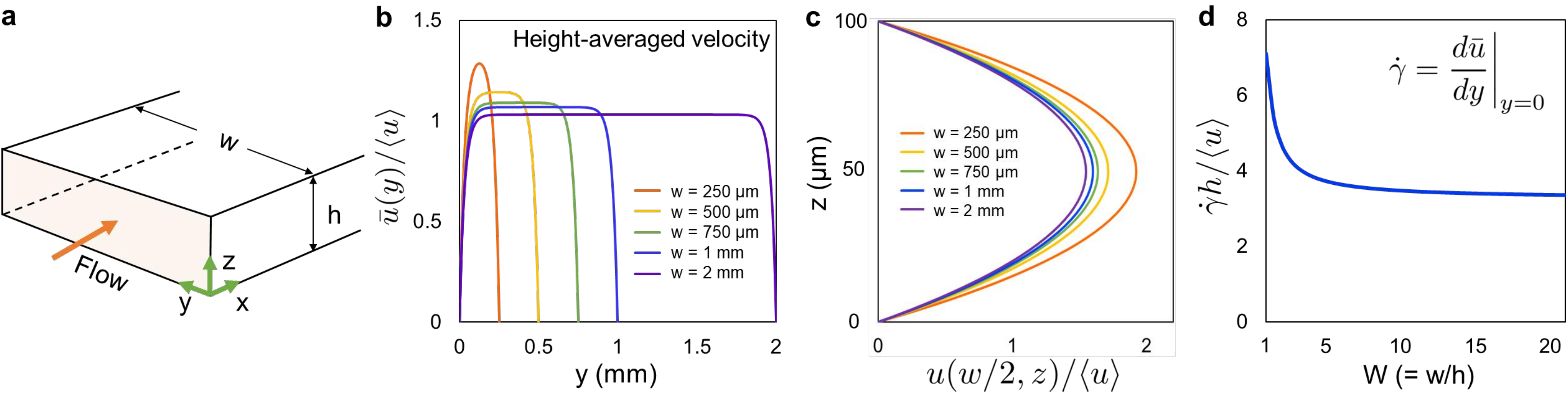}
\caption{\label{fig5} Calculation of flow velocities and shear rates for different channel dimensions. (a) Schematic of the rectangular channel. (b) Height-averaged velocity profiles ($\bar{u}(y)/\langle u\rangle$) plotted versus $y$, for different values of the channel width. (c) The height direction flow velocity normalized by the mean flow speed plotted versus channel height (note the rotated orientation of the graph). For same mean flow speeds, the centerline ($y=w/2$) velocity varies with the width of the channel. (d) Nondimensional shear rate ($\dot\gamma h/\langle u\rangle$) at the wall ($y=0$) plotted versus the aspect ratio $\mathcal{W}$. For the same mean flow speed $\langle u\rangle$ and height $h$, we can obtain varying shear rates at the wall by changing the width of the channels. }
\end{figure*}

In our model calculations, $\eta_c$ provides a direct measure for the exclusion zone profile by its definition. From the definition of the similarity variable, we note that
\begin{equation}
\eta_c = \frac{Y_c}{(3 X_c)^{\frac{1}{3}}} = \frac{y_c}{\left( 3 x_c \frac{D_c}{\dot\gamma}\right)^{\frac{1}{3}}}~,
\end{equation}
where $(x_c,y_c)$ and $(X_c, Y_c)$ are, respectively, the dimensional and nondimensional coordinates of the boundary of the particle excluzion zone. Therefore, we can write the shape of the boundary in $x,y$ coordinates as

\begin{equation}
y = \alpha \eta_c (3x (D_c/\dot\gamma))^{\frac{1}{3}}	~, \label{eq22}
\end{equation}
where $\alpha$ is a prefactor that reflects the details of the three-dimensional flow. 

With the experiment presented in Figure \ref{fig2}, we compare the steady-state exclusion zone profile with the equation (\ref{eq22}) in Figure \ref{fig4-5}. In section 4, we used the small $X$ condition $\displaystyle{\frac{k_f}{\dot\gamma}X^{\frac{2}{3}}} \ll 1$ to obtain equation (\ref{eqnfceta}). For $k_f = 0.039$ s$^{-1}$ and $\dot\gamma \approx 2$ s$^{-1}$ the condition becomes $x \ll 1.2$~cm, and thus we compare the experimental data with the equation (\ref{eq22}) up to $x=5$ mm. The wall shear rate values were estimated based on the height-averaged velocity (presented in the next section) in a channel with the width and height, respectively, $w=1$ mm and $h=100~\mu$m. In Figure \ref{fig4-5}, the curves for equation (22) represent the calculations with wall shear rate values $\dot\gamma=1.74,~3.47,~8.68,~$ and $17.37$ s$^{-1}$, corresponding to the mean flow speeds $\langle u \rangle = 50,~100,~250,~500~\mu$m/s. The steady-state exclusion zone match the scaling equation (\ref{eq22}) well with the choice of $\alpha=0.45$. Different channel geometries may require different fitting parameters. We also obtained the exclusion zone boundary from three-dimensional (3D) model calculation (details are included in the SI sections 6 and 7); and the scaling analysis, experiments and a 3D model show good agreement near the inlet of the channel (Figure \ref{fig4-5}). Note that the 3D calculation is not the main focus of the study so we only include it in the SI.

In the experimental situations, the value of $\eta_c$ is usually a constant if the particle zeta potential $\zeta$ and so the mobility $\Gamma_p$ are constant. Therefore, we conclude with the relation that the thickness of particle exclusion zone $y \propto \dot\gamma^{-\frac{1}{3}}$. This means that the size of the particle exclusion zone is determined by the wall shear rate. The trend $y \propto \dot\gamma^{-\frac{1}{3}}$ was recognized by Lee \textit{et al}.\cite{lee} ($\delta \propto (U/w)^{-\frac{1}{3}}$; $\delta$ is the exclusion zone thickness and $U$ is the mean flow speed) in the numerical simulations. In our article, we further study the rectangular channel flow to calculate the wall shear rate, and examine the effect of channel geometry and validity of the power law under multiple conditions.

So far we have confirmed with the experiments and the theoretical model based on a shear flow approximation that the flow speeds control the exclusion zone thickness. However, changing flow speed (or the volumetric flow rate) also changes the number of particles per volume that are excluded from the boundary. Therefore, in terms of ``water cleaning'' experiment, changing the flow rate to test the effect of shear rate on particle exclusion is not a single-parameter control. Thus in the next section, we study the details of flow structure in a rectangular microfluidic channel, and show how the wall shear rate $\dot\gamma$ can be varied in such a system without changing the flow speed. In that way, we can test only the effect of shear rate on the size of the particle exclusion zone, without perturbing the number of excluded particles per volume.

\smallskip
\section{Wall shear rate in a rectangular channel}

\begin{figure*}[t!]
\centering
\includegraphics[width=5.55in]{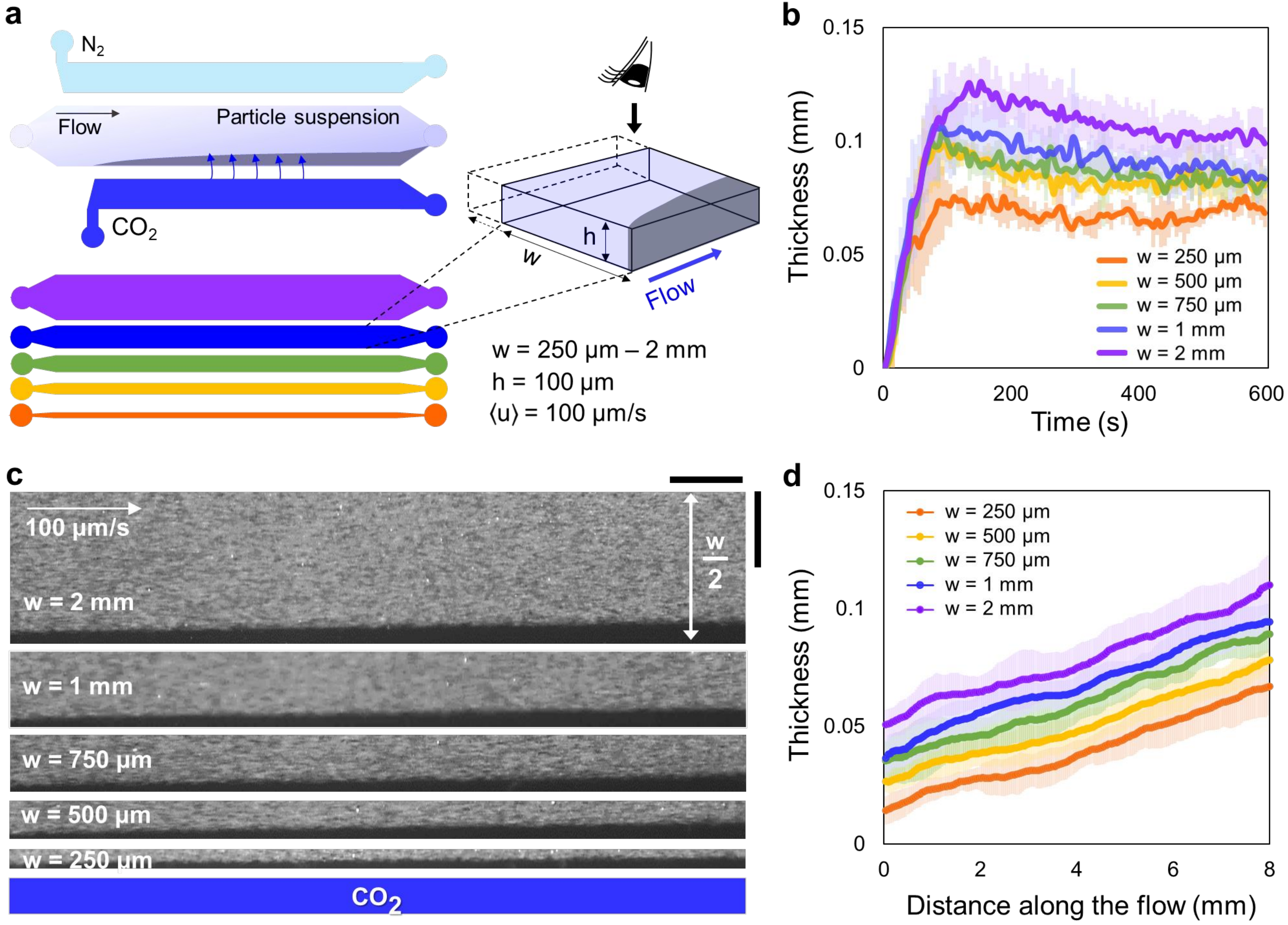}
\caption{\label{fig6} Effect of channel width on the size of the exclusion zone. (a) Schematic for the width control experiments. Channels with different widths ($w = 0.25,~0.5,~0.75,~1$ and $2$ mm) were used to demonstrate different wall shear rates under the same mean flow speed $\langle u \rangle$. (b) Time-evolution of the exclusion zone measured at $x=8$ mm. (c) Images obtained from the channels with different widths, at $t=300$ s. For $\langle u\rangle = 100~\mu$m/s and $h=100~\mu$m, change in $w$ varies the exclusion zone thickness. Horizontal and the vertical scale bars are, respectively, 1 mm and 500 $\mu$m. (d) Exclusion zone profiles at $t=300$ s plotted versus $x$. Both (b) time-evolution at a fixed location and (d) profile at $t=300$ s show larger exclusion zone formation in the channels with larger width. }
\end{figure*}

In this section, we investigate the flow structure to find important parameters for controlling the shear rate in a rectangular channel. Consider a channel with the width, height, and length, respectively $w$, $h$, and $L$. We define the axes from a corner of the channel (Figure \ref{fig5}(a)), $x$, $y$, and $z$ in the flow, width, and height directions, so that the flow velocity $\mathbf{u}=u(y,z)\mathbf{e_x}$ can be written as 
\begin{align}
\begin{split}
u(y,z)&=\frac{6 \langle u \rangle}{h^2} \left[1-6\left(\frac{h}{w}\right)\sum_{n=0}^\infty \lambda_n^{-5} \tanh \left(\lambda_n \frac{w}{h}\right)\right]^{-1}\\ 
&\times \left[(hz-z^2)-\sum_{n=0}^\infty a_n \cos \left(\frac{\lambda_n \left(z-\frac{h}{2}\right)}{h/2}\right) \cosh\left(\frac{\lambda_n \left(y-\frac{w}{2}\right)}{h/2}\right)\right], \label{uyz} 
\end{split}
\end{align} 
where $\displaystyle{a_n=\frac{h^2 (-1)^n}{\lambda_n^3 \cosh(\lambda_n w/h)}}$, $\displaystyle{\lambda_n=\frac{2n+1}{2} \pi}$, and $\langle u \rangle$ is the mean flow speed. Note that the series solution is first obtained for the coordinate system defined along the centerline of the channel, then shifted by $y=w/2$ and $z=h/2$ (see SI for details).

In the experiments, the exclusion zone forms from one side wall, which means that we are observing the height-averaged view of the system. Therefore, we consider the height-averaged velocity profile to obtain the wall shear rate at $y=0$ and simplify the analysis to a two-dimensional flow. The height-averaged velocity $\bar{u}(y)$ is (see SI for details)
\begin{equation*}
\bar{u}(y) = \frac{1}{h}\int_{0}^{h} u ~dz ~~~~~~~~~~~~~~~~~~~~~~~~~~~~~~~~~~~~~~~~~~~~~~~~~~~~~~~~~~~~~~~~~~~~~~~~~~~~
\end{equation*}
\vspace{-5ex}
\begin{align}
\begin{split}
= \langle u\rangle &\left[1- \left(\frac{6}{\mathcal{W}}\right)\sum_{n=0}^\infty \lambda_n^{-5} \tanh \left(\lambda_n \mathcal{W}\right)\right]^{-1}\left[1-\sum_{n=0}^\infty \frac{6 \cosh \left( \frac{\lambda_n \left(y-\frac{w}{2}\right)}{h/2}\right)}{\lambda_n^4 \cosh \left(\lambda_n \mathcal{W}\right)}\right]~, \label{uy}
\end{split}
\end{align}
where $\mathcal{W} = w/h$. The equation (\ref{uy}) is normalized by $\langle u\rangle$ and calculated for different values of channel widths ($w=0.25, 0.5, 0.75$ and $2$ mm) and $h=100~\mu$m, and plotted versus $y$ in Figure \ref{fig5}(b). Also, the velocity profile viewed from the side wall, along the centerline, ($u(y=\frac{w}{2},z)/\langle u\rangle$) is calculated and plotted for channels with different widths (Figure \ref{fig5}(c)). For $w \gg h$, the flow velocity is almost uniform across the width of the channel, except for the small region near the walls.

Next we calculate the shear rate $\dot\gamma$ at the wall ($y=0$ or $y=w$) using the height-averaged velocity $\bar{u}$:
\begin{equation}
\begin{aligned}
\dot\gamma &= \frac{d \bar{u}(y)}{d y}\bigg|_{y=0,~y=w} \\&= \pm\frac{12 \langle u \rangle}{h}  \left[  1- \left(\frac{6}{\mathcal{W}}\right) \sum_{n=0}^\infty \lambda_n^{-5} \tanh \left( \lambda_n \mathcal{W}\right)\right]^{-1}\left[ \sum_{n=0}^{\infty} \lambda_n^{-3} \tanh{(\lambda_n \mathcal{W})} \right]~. \label{gammadot}
\end{aligned}
\end{equation}
Note that if we maintain the mean flow speed and the channel height constant, the aspect ratio of the channel $\mathcal{W}$ is the main parameter that can vary the shear rate at the wall. The normalized shear rate $\dot\gamma h /\langle u \rangle$ is plotted versus $\mathcal{W}$ in Figure \ref{fig5}(d). This calculation shows that for the same mean flow velocity and the channel height, we can tune the wall shear rate $\dot\gamma$ simply by changing the width of the channel. For smaller $\mathcal{W}$, the shear rate is higher. We also note that (SI) the shear rates at $z=0$ and $z=h$ are smaller for larger $\mathcal{W}$, which means that for the same mean flow velocity and $h$, we obtain lower shear rates on all four walls in the channels with larger widths.

\bigskip
\section{Exclusion zone formation in the channels with various geometries}

\begin{figure*}[t!]
\centering
\includegraphics[width=7.3in]{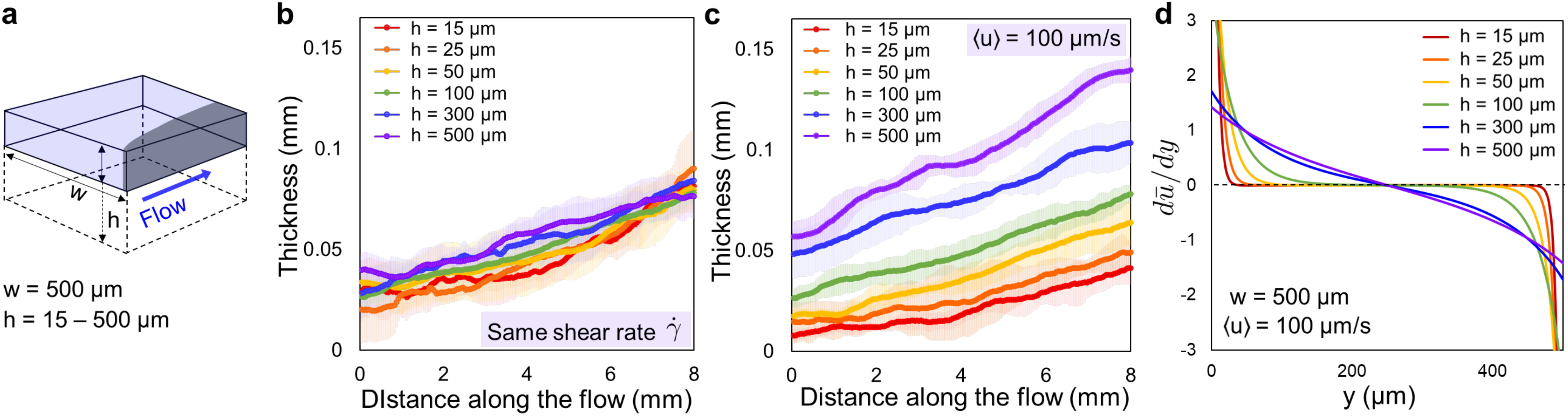}
\caption{\label{fig7} (a) Schematic for the height control experiments. Channels with the width $w=500~\mu$m and various heights were used. Exclusion zone profiles measured at $t=300$ s in the channels with different heights, (b) for the same shear rate $\dot\gamma$ and (c) for the same mean flow speed $\langle u \rangle = 100~\mu$m/s. (d) Shear rate $d\bar{u}/dy$ plotted versus $y$ for the channels with different heights. }
\end{figure*}

From the previous sections, we obtained the relation for the thickness of exclusion zone ($y\sim \dot\gamma^{-\frac{1}{3}}$) and calculated the wall shear rate $\dot\gamma$ as a function of channel dimensions. Motivated by the shear rate calculations, we performed control experiments with the rectangular channels with the height $h = 100~\mu$m and different widths $w = 0.25,~0.5,~0.75,~1$ and $2$ mm (Figure \ref{fig6}(a)). We observe the particle exclusion zone near the CO$_2$-side boundary at a fixed mean flow speed $\langle u \rangle = 100~\mu$m/s. CO$_2$ and N$_2$ pressures were maintained at, respectively, 10 psi and 1.5 psi. 300 seconds after the CO$_2$ flow is on, we obtain different exclusion zones in the channels of different widths (Figure \ref{fig6}(b-d)). 

As predicted from the calculations, the shear rate is smaller for larger $\mathcal{W}$ channels with fixed values of $\langle u\rangle$ and $h$, and thus we observe larger exclusion zones in the channels with larger width. We measured the time evolution of particle exclusion zone at $x=8$ mm and plotted versus time in Figure \ref{fig6}(b). The experiments show that the early shape of the particle exclusion zone also follows the trend predicted by the shear rate calculations, where we achieve larger exclusion zones in the channels with larger width. Next we plotted versus $x$ the shape of particle exclusion zone at $t=300$ s for different channels (distance along the channel) in Figure \ref{fig6}(d). Even considering some overlapping error bars, we observe a clear trend that the particle exclusion zone is larger in channels of larger width.

We do not directly compare this measurement with equation (\ref{eq22}) since the channel geometries are different and the estimation of $\alpha$ -- the prefactor of equation (\ref{eq22}) -- based on the experiments done in different systems may introduce additional errors. Also, the flow structure is complicated in the wall region, and this complication is quantitatively different for channels with different dimensions. Therefore, instead, we further examine the effect of shear rate experimentally by using the channels with different heights. From the equation (\ref{gammadot}) we note that the two-dimensional wall shear rate is a function of $\langle u \rangle$, $h$ and the aspect ratio $\mathcal{W}$. 

In the next set of experiments, we fix the width of the channel $w$ and vary the height $h$ and the flow speed $\langle u\rangle$. First, experiments were performed in channels of different heights and with the adjusted flow speeds to achieve fixed value of $\dot\gamma$. For a wide range of the aspect ratio ($1< \mathcal{W} <33.3$), we obtained similar exclusion zones for a single value of $\dot\gamma$. Second, experiments were performed fixing the mean flow speed $\langle u \rangle$ while varying $h$: we obtain the largest exclusion zone in the channel with $h=500~\mu$m, and the smallest for $h=15~\mu$m, which agrees with the trend predicted by the corresponding shear rates. In Lee \textit{et al.}\cite{lee}, the excluxion zone was estimated with the scaling relation $\delta\propto (\langle u\rangle/w)^{-1/3}$. We show in Figure \ref{fig7}(c) six different systems with the same values of $\langle u \rangle/w$, which demonstrate that simple estimation of the shear rate by $\langle u\rangle/w$ may not provide sufficient information about the two-dimensional exclusion zone forming in channels with different dimensions. Flow structure and the channel geometry need to be fully considered. 



\begin{figure*}[t!]
\centering
\includegraphics[width=7.36in]{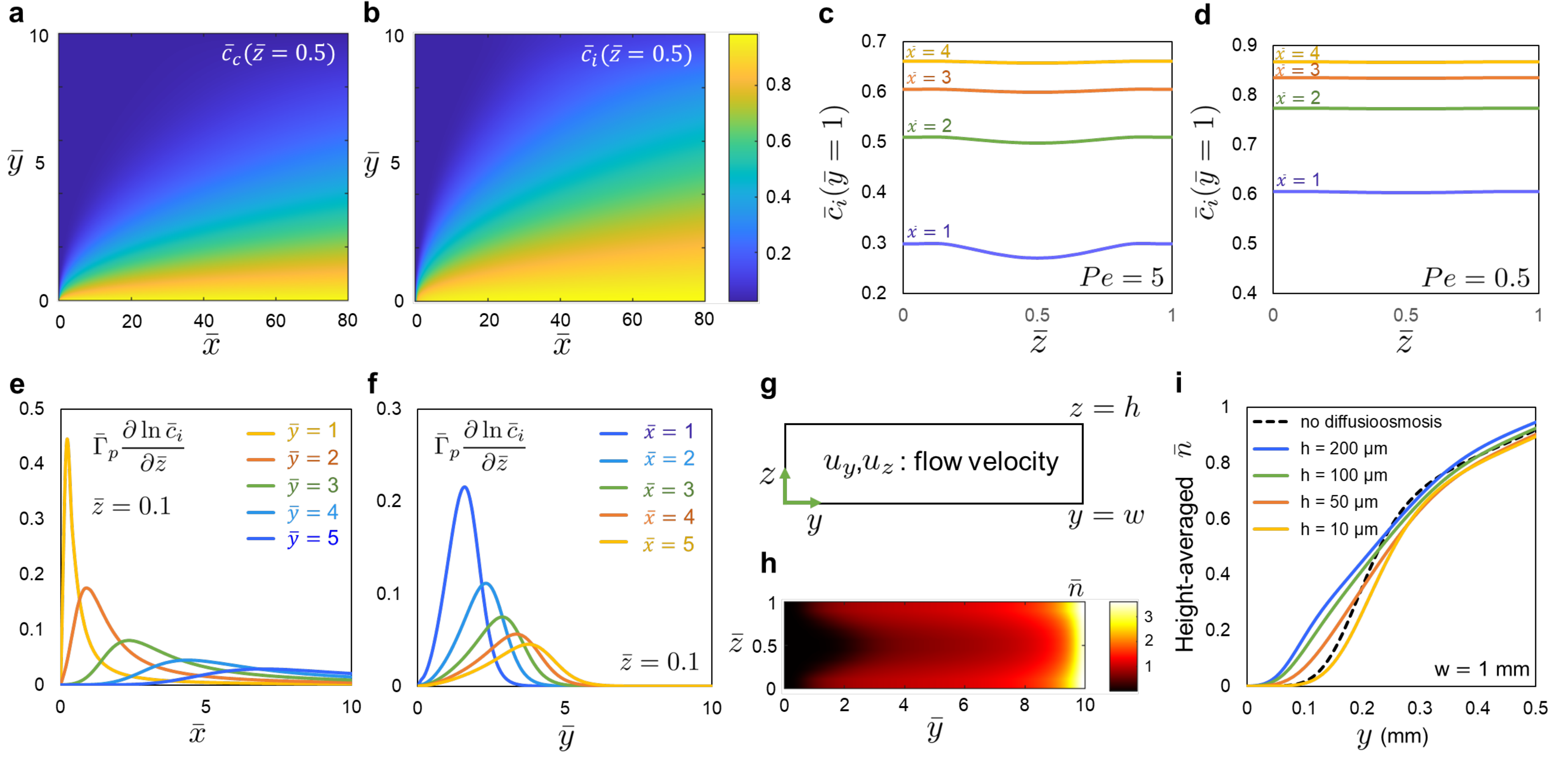}
\caption{\label{fig8} (a,b) Nondimensional, steady-state concentration profiles of CO$_2$ and ions at $\bar{z}=z/h=0.5$ plotted versus $\bar{x}=x/h$ and $\bar{y}=y/h$. (c,d) Concentration of ions at $\bar{y}=1$ plotted versus $\bar{z}$ for different Peclet numbers (note that $\mathcal{W}=10$). When diffusion is dominant, the ion concentration is more uniform along $\bar{z}$. (e,f) $\bar{z}$-component of the particle diffusiophoretic velocity plotted versus $\bar{x}$ and $\bar{y}$. Near the upstream entrance of the CO$_2$ source, highest $\bar{z}$-direction velocity is created. (g-i) Examining the effect of wall diffusioosmosis under no channel flow condition. (g) Schematic of the cross-section of the channel. We solve for the case where $u=0$, and the nonzero flow velocities $u_y$ and $u_z$ are induced by wall diffusioosmosis. (h) Nondimensional, steady-state particle concentration profile plotted versus $\bar{y}$ and $\bar{z}$. Due to the diffusioosmosis, particle distribution is not uniform along the $\bar{z}$-direction. (i) Height-averaged $\bar{n}$ plotted versus dimensional $y$, for different channel height values. The profiles are obtained at $t=300$ s.}
\end{figure*}

We would like to mention more about shear flow approximation in microfluidic experiments. Figure \ref{fig5}(d) shows $d \bar{u}/dy$ plotted versus $y$. When the aspect ratio of the channel is large, the change in the height-averaged flow speed is mostly localized near the wall region, and for small $w/h$, the change in the shear rate is less dramatic. Therefore, a typical experimental setup for testing the shear flow approximation is in the orientation where the observation is made through the small-width window ($\mathcal{W}<1$) by making tall channels. In that way the experiments can assume a constant shear rate in the wall region. Knowing this, one can ask about the effect of the orientation of the channel (whether $\mathcal{W}\geq1$ or $\mathcal{W}\leq1$)) on the water cleaning efficiency, under a fixed volumetric flow rate.

Similar to the height-averaged velocity $\bar{u}(y)=\frac{1}{h}\int_0^h u ~dz$, we calculated the width-averaged velocity $\tilde{u}(z)=\frac{1}{w}\int_0^w u~dy$ to obtain the equivalent wall shear rate in the other orientation in the same channel $\dot\gamma_z=\frac{d\tilde{u}}{d z}\big|_{z=0,~z=h}$ (see SI section 5 for details). When $\dot\gamma$ and $\dot\gamma_z$ are compared (Figure S5), we obtain that the two-dimensional shear rate is always larger when the system is viewed from the narrow-side. For example, in a channel with $w=500$ $\mu$m and $h=100$ $\mu$m, for $\langle u \rangle = 100~\mu$m/s, we obtain $\dot\gamma = 3.73$ and $\dot\gamma_z = 6.12$. When the ion concentration gradient is created over $w$, the measured exclusion zone is expected to be $(\frac{3.73}{6.12})^{-\frac{1}{3}}\approx1.2$ times larger than the other orientation. However, the height is 5 times larger if the ion concentration gradient is formed across $h$, and thus the volume of particle-free water is $\approx 4.2$ times larger. An optimization is required for specific design and fabrication of an actual water cleaning device. Nevertheless, such shear rate criterion can be a major reference for designing the diffusiophoresis unit.

So far we have used a shear flow approximation, and, with similarity transform, we obtained a scaling estimate for the exclusion zone thickness along the flow. The relation $y\propto \dot\gamma^{-\frac{1}{3}}$, along with the shear rate calculation $\dot\gamma = f(\langle u\rangle, h, \mathcal{W})$, provides useful information for predicting the trend of particle exclusion in two-dimensional configurations. In the next section, with numerical calculations we would like to investigate possible three-dimensional effects that can influence the exclusion zone measurements.

\smallskip
\section{Three-dimensional effects and wall diffusioosmosis}

Consider a three-dimensional rectangular flow (Figure \ref{fig5}(a)) with the flow velocity defined as equation (\ref{uyz}). Then we can solve reaction-diffusion equations to obtain the CO$_2$ and ion concentrations $c_c(x,y,z,t)$ and $c_i(x,y,z,t)$ (details in SI sections 6 and 7).

As a model case, we consider a channel flow with $w=1$ mm, $h=100~\mu$m, and $\langle u \rangle = 100~\mu$m/s, and plotted the steady-state concentration profiles of CO$_2$ and ions in Figure \ref{fig8}(a,b) versus $\bar{x}=x/h$ and $\bar{y}=y/x$ (at $\bar{z}=z/h=0.5$). In this configuration, it is often assumed that the concentration profiles are independent of ${z}$. However, we note that due to the finite diffusivities of ions, there is a non-uniform concentration distribution of ions in the ${z}$-direction (Figure \ref{fig8}(c)), near the upstream of the CO$_2$ source. When the system is diffusion-dominated (Figure \ref{fig8}(d)), the ${z}$-dependence of ion concentration is small. This result suggests that even though the main concentration gradient is established in ${y}$-direction, there may be a nonzero diffusiophoretic velocity of particles generated in the height-direction, affecting the particle distribution. 

The $\bar{z}$-component of nondimensional diffusiophoretic velocity $\bar{u}_{pz}=\bar{\Gamma}_p\frac{\partial \ln \bar{c}_i}{\partial \bar{z}}$ is calculated numerically near the bottom wall ($\bar{z}=0.1$) and plotted versus $\bar{x}$ and $\bar{y}$ in Figure \ref{fig8}(e,f). Near the inlet of the flow there are nonzero particle velocities forming in $z$-direction to migrate particles away from the wall ($\bar{z}=0$). Then for $\bar{x} \gg 1$ this particle velocity decreases dramatically. A similar trend is observed also in the width direction; the $\bar{z}$-component of the diffusiophoretic velocity is highest near the wall where CO$_2$ diffuses in. The nondimensional particle velocity is a measure for the particle Peclet number by definition $Pe_p = u_p h/D_c$ and the calculated values suggest that the $\bar{z}$-direction particle motion is much smaller compared to the main flow speed ($Pe = 5$). Since the ${z}$-direction gradient of ion concentration is generated due to the channel flow and the finite diffusivity of ions, larger $\bar{u}_{pz}$ will be obtained for faster flow, and for slow flow $\bar{u}_{pz}$ will be negligible. This means that the particle exclusion from the top and bottom walls is expected to be smaller compared to the main exclusion zone formation in the ${y}$-direction in the presence of the channel flow. A study of the detailed three-dimensional particle distribution is not the main focus of the current article and so we include the contour plot of particle exclusion in the SI. The feature discussed in this section may be of consideration for maximizing the collection of clean liquid in the in-flow water cleaning systems. 

\begin{figure*}[t!]
\centering
\includegraphics[width=7.36in]{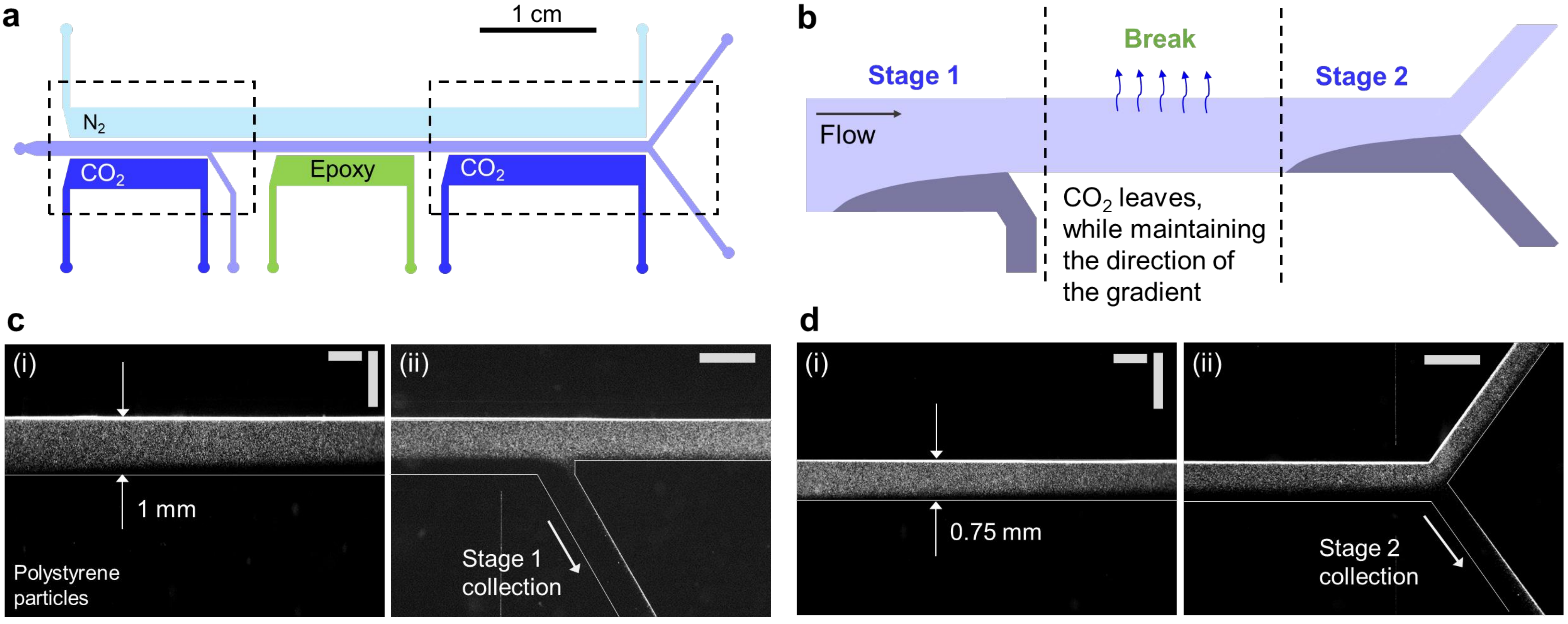}
\caption{\label{fig9} Two-stage water cleaning system. (a,b) Schematics of the channel geometry and the experiment. (c) Clean water collection in the first stage. (d) Clean water and particle suspension collection at the end of the second stage. Scale bars are 1 mm. }
\end{figure*}

One more feature to study in terms of a three-dimensional effect is the diffusioosmosis on the (top and bottom) walls. When there is a pressure-driven flow with a moderate speed, analyses that consider particle diffusiophoresis and the flow velocity may provide sufficient information about the exclusion zone formation, even though there exists diffusioosmosis forming a slip boundary condition. We consider one extreme case where the flow velocity is very small (or zero) and the exclusion zone formation is achieved by mostly (or only) the diffusion of ions. Small flow velocity decreases the overall efficiency of the water cleaning system, but if the flow is deliberately stopped periodically to achieve maximum exclusion zone thickness before each sample collection step (pulsating flow), one may ask about the influence of the wall diffusioosmosis. 

We consider a model case where there is no flow along the channel and the migration of particles in the channel is driven by diffusion of ions. We can solve for the cross-sectional diffusion of ions and diffusiophoresis of particles and examine the effect of diffusioosmosis (Figure \ref{fig8}(g)). First, we account for diffusioosmosis along the top and bottom walls, to obtain the flow velocity ($\mathbf{u}_f=u_y \mathbf{e}_y+u_z \mathbf{e}_z$) as a function of $c_i$ (more precisely, we use the local chemical equilibrium approximation \cite{hsc,orest}, and use $c_c$ to describe the flow velocity induced by wall diffusioosmosis; see SI section 8 for details).

In Figure \ref{fig8}(h), the steady-state particle distribution $\bar{n}(\bar{y},\bar{z})$ is plotted versus $\bar{y}=y/h$ and $\bar{z}=z/h$. Due to the flow velocity induced by the wall diffusioosmosis, particle distribution is not uniform along $\bar{z}$ under the concentration gradient set in the $\bar{y}$-direction. We solved for the particle distribution for the channel cross-sections with different aspect ratios and plotted the $\bar{z}$-averaged profiles versus dimensional $y$ (Figure \ref{fig8}(i)). For the channel with $w=1$ mm, the $\bar{z}$-averaged $\bar{n}$ at $t=300$ s is plotted for different channel heights. The plots show that at smaller heights, the effect of diffusioosmosis is larger and exclude particles more. Such difference in particle exclusion may be significant in microscopic analyses of diffusiophoretic particle exclusion. However, the slight difference among different conditions would not affect practical applications, and the comparison with the no-diffusioosmosis scenario also suggests that the effect of wall diffusioosmosis can be neglected.

Throughout the paper, we showed by a theoretical model, experiments and numerical calculations that the thickness of particle exclusion zone near entrance follows the trend $y \propto \dot\gamma^{-\frac{1}{3}}$, where the shear rate varies with the channel dimensions. By increasing the width and height of the rectangular channel, we were able to achieve formation of larger exclusion zone. With these understandings, we would like to propose a possible two-step design for the in-flow water cleaning system (Figure \ref{fig9}). Again, the design suggestion is not an optimal geometry or an ultimate solution. We chose conventional microfluidic fabrication to visualize the concept in a small length scale. The ideas should be further developed and customized to a real device application.  

\bigskip
\section{Toward multi-stage water cleaning}

In order to visualize the idea of two-stage water cleaning, we fabricated a PDMS channel ($w_1=1$ mm and $h=100~\mu$m) that has the first bifurcation $\approx 1.5$ cm downstream from the inlet, and the second bifurcation $\approx 5.5$ cm downstream from the inlet. After the first bifurcation, the width of the remaining liquid channel is $w_2=750~\mu$m. In order to effectively create the CO$_2$ (or ion) concentration gradients in the second stage, we put a ``break'' stage where the liquid channel is in contact with the N$_2$ channel and a channel filled with the cured optical adhesive (Norland Products; NOA 81 -- Figure \ref{fig9}(a,b)). In the break stage, CO$_2$ that is in the liquid phase is expected to diffuse out on the side facing the N$_2$ channel, as the UV epoxy is impermeable to gas. While maintaining the direction of the ion concentration gradient, the absolute concentration of ions decrease at this stage\cite{pulse} (SI). In this way, we do not perturb the particle distribution by introducing backward ion concentration gradients. Then in the second stage, an exclusion zone is formed on the same side as the first stage, and the clean water can be collected at the final bifurcation outlet. In order to match the size of the exclusion zones with the model channel design, we adjusted the flow speed to a lower value ($\langle u \rangle = 10~\mu$m/s) than our main experiments (Figure \ref{fig9}(c,d)).

\bigskip
\section{Conclusion}
Our study on the diffusiophoretically-generated exclusion zone is motivated by several previous studies that demonstrate diffusiophoresis in microfluidic systems. By increasing the width of the channels to $O(1)$ mm, we observed that the exclusion zone forms in a small region near the wall facing the CO$_2$ source. Then, with the shear flow approximation and similarity transform, we obtained a scaling law for the exclusion zone thickness ($y \propto \dot\gamma^{-\frac{1}{3}}$), which led to a detailed calculation for the shear rate. The wall shear rate in a rectangular channel flow is a function of $\langle u \rangle$, $\mathcal{W}$ and height $h$, so the channel geometry and the flow condition can be chosen to obtain a desired wall shear rate. With control experiments we show that the exclusion zone prediction using the shear rate is applicable to channels with various widths and heights. We also confirmed experimentally that a change in the applied CO$_2$ pressure does not significantly vary the exclusion zone thickness in our system. Finally, we propose a design to demonstrate a continuous, multi-stage water cleaning. As the most water purification process is never a single step, by an appropriate choice of the geometrical parameters, such in-flow water cleaning system can be further scaled up to collect larger amount of cean water. 

\bigskip
\section*{Contributions}
S.S. and H.A.S. conceived the project. S.S. designed all experiments with assist of M.B., and S.S. M.B., and E.H.T. performed the experiments. S.S. and H.A.S. set up the theoretical model, and S.S. conducted numerical calculations. M.B. and E.H.T. contributed to the project through the Independent Work and Princeton Environmental Institute Summer Internship Program during their undergraduate study at Princeton University. 

\bigskip
\section*{Conflicts of interest}
S. Shim declares no conflict of interest. H. A. Stone co-founded Phoresis, Inc., which seeks to utilize CO$_2$-driven diffusiophoresis in the water-cleaning space.

\bigskip
\section*{Acknowledgements}
We acknowledge NSF for support via CBET-1702693. We thank Jesse T. Ault and Estella Yu for valuable discussions. M.B. and E.H.T. thank the Princeton Environmental Institute at Princeton University for supporting summer internships.

\bigskip
\section*{Materials and Methods}

The channels are made with PDMS using standard soft lithography. All four walls are PDMS, and the bottom layer (1-mm PDMS slab) is further attached to the glass slide for microscopy (Leica DMI4000B). For all experiments using polystyrene (PS) particles (Thermo Fisher Scientific, diameter = 1~$\mu$m), a dilute suspension of 0.03 vol$\%$ (in DI water; Milli-Q, EMD Millipore) is used. 

For the experiment with amine-modified polystyrene (a-PS) particles (Fig. \ref{fig1}(d); Sigma Aldrich, diameter 1 $\mu$m), a dilute suspension is prepared at 5 vol$\%$ in DI water. Prior to the experiment, a syringe pump was used to flow a 1 vol$\%$ aqueous solution of 3-Aminopropyltriethoxysilane (APTES) through the liquid channel for 15 minutes to prevent the adhesion of amine-modified polystyrene particles to channel walls \cite{PDMSmod1,PDMSmod2}, followed by flushing with deionized water for 15 minutes. A N$_2$ gas stream at 1.5 psi is connected for one minute to dry the central channel. 

When running the experiments, the liquid channel was first filled with the particle suspension, then N$_2$ gas was flowed at 1.5 psi through the top channel. After one minute of observing steady-state particle flow in the liquid channel, a CO$_2$ gas stream fixed at 10 psi is connected to the CO$_2$ channel.  

The fluorescent images are obtained every 1 second using an inverted microscope (Leica DMI4000B), with 1.25x magnification. The exclusion zone boundary was obtained by first dividing the images into 200 separate sectionst (in $x$-direction), then detecting the mean value of the boundary position in each section. Each image piece had the width of 6 pixels, and thus the representative data point is an average of six values. The obtained values ($y$-position of the exclusion zone boundary) are further smoothed by applying a moving average of period 20. For the time-evolution plots, a similar method is applied, where the mean value of the boundary position is obtained every six frames. No smoothing is applied for the time evolution graphs. The smooth curves in Fig. 1(e) and Fig. S1(b) are obtained using a smoothing spline fit in Matlab. The image processing and analysis are done with ImageJ and Matlab.  

All equations presented in the main text are solved using a combination of Mathematica and Matlab, and the three-dimensional model included in the SI is solved with Matlab (see SI for details).



\balance


\bibliography{rsc} 
\bibliographystyle{rsc} 

\end{document}